\documentclass[fleqn,12pt,twoside]{article}
\usepackage{espcrc1}
\usepackage{wrapfig}

\usepackage{graphicx}
\usepackage[figuresright]{rotating}

\newcommand{\AmS}{{\protect\the\textfont2
  A\kern-.1667em\lower.5ex\hbox{M}\kern-.125emS}}

\title{Nucleon-nucleon coincidence measurement in the non-mesonic weak
decay of $^{5}_{\it{\Lambda}}$He and $^{12}_{\it{\Lambda}}$C
hypernuclei}

\author
{ S.~Okada$^{a,}$\footnote{Present address: RIKEN Wako Institute, RIKEN,
  Wako 351-0198, Japan},
  S.~Ajimura$^b$,
  K.~Aoki$^c$,
  A.~Banu$^d$,
  H.~C.~Bhang$^e$,
  T.~Fukuda$^{c,}$\footnote{Present address: Laboratory of Physics,
  Osaka Electro-Communication University,
  572-8530, Japan},
  O.~Hashimoto$^f$,  
  J.~I.~Hwang$^e$,
  S.~Kameoka$^f$,  
  B.~H.~Kang$^e$,
  E.~H.~Kim$^e$,
  J.~H.~Kim$^{e,}$,
  M.~J.~Kim$^e$,
  T.~Maruta$^g$,
  Y.~Miura$^f$,
  Y.~Miyake$^b$,
  T.~Nagae$^c$,
  M.~Nakamura$^g$,  
  S.~N.~Nakamura$^f$,
  H.~Noumi$^c$,
  Y.~Okayasu$^f$,
  H.~Outa$^{c,*}$,
  H.~Park$^h$,
  P.~K.~Saha$^{c,}$\footnote{Present address: Japan Atomic Energy
  Research Institute, Tokai 319-1195, Japan},
  Y.~Sato$^c$,
  M.~Sekimoto$^c$,
  T.~Takahashi$^{f,}$\footnote{Present address: High Energy Accelerator
  Research Organization (KEK), Tsukuba 305-0801, Japan},
  H.~Tamura$^f$,
  K.~Tanida$^i$,
  A.~Toyoda$^c$,
  K.~Tsukada$^f$,
  T.~Watanabe$^f$,
  H.~J.~Yim$^e$ \\
  \vspace{2mm}
  $^a$Department of Physics, Tokyo Institute of Technology,
  Ookayama 152-8551, Japan \\
  $^b$Department of Physics, Osaka University,
  Toyonaka 560-0043, Japan \\
  $^c$High Energy Accelerator Research Organization (KEK),
  Tsukuba 305-0801, Japan \\
  $^d$Gesellschaft f$\ddot{\mbox{u}}$r Schwerionenforschung mbH (GSI),
  Darmstadt 64291, Germany \\
  $^e$Department of Physics, Seoul National University,
  Seoul 151-742, Korea \\
  $^f$Department of Physics, Tohoku University,
  Sendai 980-8578, Japan \\
  $^g$Department of Physics, University of Tokyo,
  Hongo 113-0033, Japan \\
  $^h$Korea Research Institute of Standards and Science (KRISS),
  Daejeon 305-600, Korea \\
  $^i$RIKEN Wako Institute, RIKEN,
  Wako 351-0198, Japan \\
  }
       
\begin{document}

\maketitle

\begin{abstract}

 We have measured both yields of
 neutron-proton and neutron-neutron pairs
 emitted from the non-mesonic weak decay
 process of $^{5}_{\it{\Lambda}}$He and $^{12}_{\it{\Lambda}}$C
 hypernuclei produced via the ($\pi^+$,$K^+$) reaction
 for the first time.
 We observed clean back-to-back correlation of the $np$- and $nn$-pairs
 in the coincidence spectra for both hypernuclei.
 The ratio of those back-to-back pair yields,
 $N_{nn} / N_{np}$,
 must be close to the ratio of neutron- and proton-induced
 decay widths of the decay,
 $\Gamma_n$($\it{\Lambda}n \to nn$)/$\Gamma_p$($\it{\Lambda}p \to np$).
 The obtained ratios for each hypernuclei
 support recent calculations based on short-range interactions.

\end{abstract}

\section{Introduction}

The non-mesonic weak decay (NMWD) process
of a $\it{\Lambda}$ hypernucleus, $\it{\Lambda} N \to nN$,
gives a unique opportunity to study the weak interaction
between baryons
since this strangeness non-conserving process is purely attributed to
the weak interaction.
In the NMWD, there are two decay channels,
$\it{\Lambda}p$ $\to n p$ ($\Gamma_p$) and
$\it{\Lambda}n$ $\to n n$ ($\Gamma_n$).
The ratio of those decay widths,
$\Gamma_n$/$\Gamma_p$,
is an important observable
used to study the isospin structure of the NMWD mechanism.
For the past 40 years, there has been a longstanding puzzle
that the experimental $\Gamma_n$/$\Gamma_p$ ratio
disagrees with that of theoretical calculations
based on the most natural and simplest model,
the One-Pion Exchange model (OPE).
In this model, the $\it{\Lambda} N \to nN$ reaction is expressed
as a pion absorption process 
after the $\it{\Lambda} \to N \pi$ decay inside the nucleus.
Since the OPE process is tensor-dominant
and the tensor transition of the initial $\it{\Lambda} N$ pair
in the $s$-state requires the final $n N$ pair to have isospin zero,
the $\Gamma_n/\Gamma_p$ ratio in the OPE process becomes close to 0.
However, previous experimental results have indicated
a large $\Gamma_n/\Gamma_p$ ratio ($\sim$1) \cite{Szy91,Noumi95}.

This large discrepancy between the OPE-model predictions and
the experimental results has stimulated many theoretical studies:
the heavy meson exchange model,
the Direct Quark model
and the two-nucleon (2$N$) induced model
($\it{\Lambda} N N \rightarrow n N N$).
After K. Sasaki $\textit{et~al.}$
pointed out an error in the sign
of the kaon exchange amplitudes in 2000 \cite{Sas00},
those theoretical values of the $\Gamma_n/\Gamma_p$ ratio
have increased to the level of 0.4$\sim$0.7 \cite{Alb02}.

On the other hand, the experimental data still have large errors
($\Gamma_n/\Gamma_p$ = 0.93 $\pm$ 0.55
for $^{5}_{\it{\Lambda}}$He \cite{Szy91}),
and it is hard to draw a definite conclusion
on the $\Gamma_n/\Gamma_p$ ratio.
When we compare the measured $\Gamma_n/\Gamma_p$ ratio with that
obtained in theoretical calculations, the most serious technical
problem was a treatment of the re-scattering effect in the residual
nucleus, the so-called Final State Interaction (FSI).
Moreover, the possible existence of a multi-nucleon induced process
has been discussed theoretically (such as 2$N$-induced process),
though there has been no experimental evidence.
Several nucleon energy spectra from hypernuclear decay
have been reported so far \cite{Sat03,Kim03},
in which it is however difficult to extract the 
$\Gamma_n/\Gamma_p$ ratio
without theoretical assumptions on the effects
of FSI and possible multi-nucleon induced processes.

Since the 1$N$-induced decay ``$\it{\Lambda} N \rightarrow n N$'' is
two-body process, the outgoing nucleon-nucleon pair suffering no FSI
effect must have about 180 degree opening angle and clear energy
correlation.  In the present experiment, we performed
a coincident measurement of the two nucleons, $np$ and $nn$-pairs,
in the decay for the first time.
The 1$N$-induced processes could be clearly observed
by measuring yields of the back-to-back $np$- and $nn$-pairs
and confirming that the energy sums 
roughly correspond to their $Q$-values ($\sim$150 MeV).
The measured yields of the coincident back-to-back $np$- and $nn$-pairs,
$Y_{np(nn)}$, are represented as
$Y_{np(nn)} = N_{np(nn)} \cdot \Omega_{np(nn)} \cdot \varepsilon_{np(nn)}
\cdot (1 - R_{FSI})_{np(nn)}$,
where $N_{np (nn)}$ are the number of back-to-back $np$($nn$)-pair events
from the decay;
$\Omega_{np (nn)}$, $\varepsilon_{np(nn)}$ and $(1-R_{FSI})_{np (nn)}$
stand for decay-counter acceptances and detection efficiencies and
reduction factors (due to the FSI or/and other non back-to-back processes)
for the $np$($nn$)-pair, respectively.
It is noteworthy that the reduction factors
are approximately canceled out
with assumption of the charge symmetry,
$(1 - R_{FSI})_{np} \simeq (1 - R_{FSI})_{nn}$,
when we take the ratio of the $np$- and $nn$-pair yields,
$N_{nn} / N_{np}$.

In order to minimize the FSI effect,
we selected a light $s$-shell hypernucleus, $^{5}_{\it{\Lambda}}$He.
In $s$-shell hypernucleus,
initial relative $\it{\Lambda} N$ states must be $S$ states,
whereas in a $p$-shell hypernucleus they may be $P$ states.
To investigate the $p$-wave effect,
we also performed the same experiment
for a typical light $p$-shell hypernucleus, $^{12}_{\it{\Lambda}}$C.

In this Letter, we show the opening angle and the energy sum distributions
of $np$- and $nn$-pairs from the NMWD of $^{5}_{\it{\Lambda}}$He and
the $N_{np}/N_{nn}$ ratio for both hypernuclei.

\section{Experimental method}

The present experiments (KEK-PS E462/E508) were carried out
at the 12-GeV proton synchrotron (PS)
in the High Energy Accelerator Research Organization (KEK).
Hypernuclei, $^{5}_{\it{\Lambda}}$He and $^{12}_{\it{\Lambda}}$C,
were produced via the ($\pi^+$,$K^+$) reaction at 1.05 GeV/$c$
on $^6$Li and $^{12}$C targets, respectively.
Since the ground state of $^{6}_{\it{\Lambda}}$Li
is above the threshold of $^{5}_{\it{\Lambda}}$He $+\ p$,
it promptly decays into
$^{5}_{\it{\Lambda}}$He emitting a low-energy proton.
The $^{6}$Li ($\pi^+$,$K^+$) $^{6}_{\it{\Lambda}}$Li reaction
was therefore employed to produce $^{5}_{\it{\Lambda}}$He.
The hypernuclear mass spectra were calculated by
reconstructing the momenta of incoming $\pi^+$ and outgoing $K^+$
using a beam-line spectrometer composed of the QQDQQ system
and the superconducting kaon spectrometer (SKS) \cite{Fuk95},
respectively.

Particles emitted from the decays of $\it{\Lambda}$ hypernuclei
were detected by the decay-particle detection system
installed symmetrically in the direction to the target
in order to maximize acceptance of the back-to-back event
for $np$- and $nn$-pairs from the NMWD process,
as shown in Ref.\cite{Oka04} (Fig.\ 1).
It was composed of plastic scintillation counters
and multi-wire drift chambers.
The decay particles were identified
by the time-of-flight and the range.

\section{Analysis and Results}

\begin{wrapfigure}[21]{r}{0.65\linewidth}
 \begin{flushleft}
  \begin{center}
   \vspace*{-19mm}
   \includegraphics[width=1.00\linewidth]{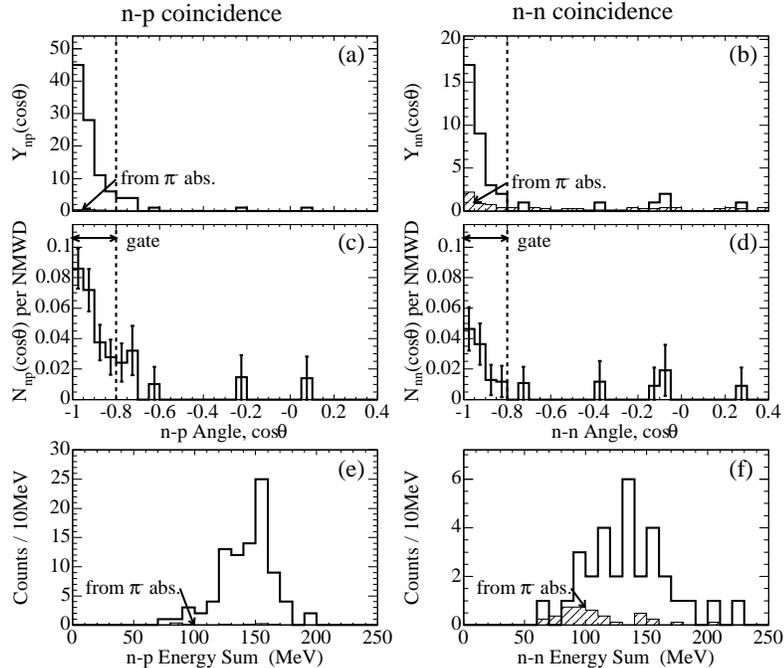}
   \vspace*{-11mm}
   \caption{{\footnotesize Upper figures show opening angle distributions
   of $np$- and $nn$-pairs emitted from the decay
   of $^{5}_{\it{\Lambda}}$He:
   (a) and (b) are for raw yields;
   (c) and (d) are for yields per NMWD.
   Lower figures, (e) and (f),
   show energy sum distributions of the $np$- and $nn$-pairs.}
   }
   \label{coinfig}
  \end{center}
 \end{flushleft}
\end{wrapfigure}

The ground state yields of 
$^{5}_{\it{\Lambda}}$He and $^{12}_{\it{\Lambda}}$C are, respectively,
about 4.6 $\times$ 10$^4$ and 6.2 $\times$ 10$^4$ events,
which were one order-of-magnitude higher
than those of previous experiments.
The inclusive excitation-energy spectra
of $^{6}_{\it{\Lambda}}$Li and $^{12}_{\it{\Lambda}}$C
are shown in Ref.\cite{Oka04} (Fig.\ 2).

For details of the neutral and charged
decay particle identification, refer to Ref.\cite{Oka04}.
In this Letter, we focus on the coincidence analysis of
$np$- and $nn$-pairs from the NMWD.

Upper figures of Fig.\ \ref{coinfig}, (a) and (b), show
opening angle distributions of $np$- and $nn$-pairs
at the energy threshold level of 30 MeV
for both of proton and neutron.
They seem to have clear back-to-back correlations,
though these are not corrected the angular dependent acceptance.
The shaded histogram shows estimated nucleon contaminations due to 
the pion absorption process
in which $\pi^-$'s from the mesonic decay of $\it{\Lambda}$ hypernucleus
are absorbed by the materials around the target.
The background was estimated
by assuming that the shape of the angular distribution
from this $\pi^-$ absorption process is the same as
that from the $\pi^-$ decay of $\it{\Lambda}$
($\it{\Lambda}$ $\to \pi^- p $) formed
via the quasi-free formation process
(see Ref.\cite{Oka04} for the detail).

The angular distributions of middle of Fig.\ \ref{coinfig},
(c) and (d), are corrected for
acceptances and efficiencies for $np$- and $nn$-pairs,
and normalized per NMWD.
The estimated contamination due to
the pion absorption stated above are subtracted.
They still have back-to-back correlation,
which indicates that the FSI effect is not so severe and 
1$N$-induced NMWD (two body process) is the major one.

Lower figures of Fig.\ \ref{coinfig}, (e) and (f), show
energy sum distributions of the $np$- and $nn$-pairs
by gating back-to-back events
as shown in the upper figures ($\cos \theta < -0.8$).
We confirmed that those energy sum distributions have broad peak
around these $Q$-values as expected.
The shaded histogram shows estimated contaminations
due to pion absorption as described above,
which distributes to lower energy region.

Also for $^{12}_{\it{\Lambda}}$C,
similar distributions of the angle and energy sum of the $np$- and $nn$-pairs
were obtained in a same way.
We successfully observed
$np$- and $nn$-pairs from the NMWD
of $^{5}_{\it{\Lambda}}$He and $^{12}_{\it{\Lambda}}$C.
The ratio of the back-to-back $np$- and $nn$-pair yields,
$N_{nn}/N_{np}$,
for $^{5}_{\it{\Lambda}}$He and $^{12}_{\it{\Lambda}}$C
were obtained as
 \begin{eqnarray}
  \Gamma_n/\Gamma_p~(\mbox{for~} ^{5}_{\it{\Lambda}}\mbox{He}) &\sim& N_{nn}/N_{np} ~~=~~
   0.45 \pm 0.11 \pm 0.03 ~~, \nonumber \\ 
  \Gamma_n/\Gamma_p~(\mbox{for~} ^{12}_{\it{\Lambda}}\mbox{C}) &\sim& N_{nn}/N_{np} ~~=~~
   0.40 \pm 0.09 \mbox{~~(preliminary)} ~~,\nonumber
 \end{eqnarray}
where the quoted systematic errors mainly come from
the neutron detection efficiency ($\sim$ 6 \%).
They can be approximately regarded as the $\Gamma_n/\Gamma_p$
with assumption of the charge symmetry.

\vspace*{5mm}

It is now revealed that the $\Gamma_n/\Gamma_p$ ratio is
significantly less than unity, thus excluding the earlier claim that
the ratio is close to unity \cite{Szy91}.
On the contrary, recent theoretical calculations
seem to be supportive to our results
being on the increase of the ratio toward 0.5.
The present results have finally given the answer to the longstanding
$\Gamma_n/\Gamma_p$ ratio puzzle, and have made a significant
contribution to the study of the NMWD.

\end{document}